\documentclass[12pt]{article}
\usepackage{latexsym}
\usepackage{amsfonts}

\catcode`\@=11

\def\section{\@startsection{section}{1}{\z@}{3.5ex plus 1ex minus
 .2ex}{2.3ex plus .2ex}{\bf}}

\def\thesubsection{\arabic{section}.\arabic{subsection}}
\renewcommand{\subsection}[1]{\addtocounter{subsection}{1}
\vspace{2.5mm}\par\noindent {\it \thesubsection . #1}\par
 \vspace{0.5mm} }
\catcode`\@=12
\newfont{\mbm}{msbm10 scaled\magstep1}

\DeclareFontFamily{U}{rsf}{} \DeclareFontShape{U}{rsf}{m}{n}{
  <5> <6> rsfs5 <7> <8> <9> rsfs7 <10-> rsfs10}{}
\DeclareMathAlphabet\Scr{U}{rsf}{m}{n}

\mathchardef\varGamma="0100
\mathchardef\varDelta="0101
\mathchardef\varTheta="0102
\mathchardef\varLambda="0103
\mathchardef\varXi="0104
\mathchardef\varPi="0105
\mathchardef\varSigma="0106
\mathchardef\varUpsilon="0107
\mathchardef\varPhi="0108
\mathchardef\varPsi="0109
\mathchardef\varOmega="010A

\def\drawbox#1#2{\hrule height#2pt\hbox{\vrule width#2pt height#1pt
 \kern#1pt\vrule width#2pt}\hrule height#2pt}

\def\Asym#1#2{\vcenter{\vbox{\drawbox{#1}{#2}\kern-#2pt\drawbox{#1}{#2}}}}

\begin{document}
\begin{titlepage}

\thispagestyle{empty}

\begin{flushright}
\hfill{CERN-PH-TH/2004-132}
\end{flushright}

\vspace{35pt}

\begin{center}{ \LARGE{\bf
Orientifolds, Brane Coordinates and Special Geometry \footnote{To
appear in the proceedings of ``DeserFest'' and ``The status of
M-theory'' Ann Arbor, Michigan, 3-6 April 04. }}}

\vspace{60pt}

{\bf R. D'Auria $^\star$, S. Ferrara $^{\dagger\ddag\sharp}$ and
M. Trigiante $^\star $}

\vspace{15pt}

$\dagger${\it  CERN, Theory Division, CH 1211 Geneva 23,
Switzerland. }\\ $\ddag${\it INFN, Laboratori Nucleari di
Frascati, Italy.}\\
$\sharp$ {\it University of California, Los Angeles, USA}

 \vspace{15pt}

$\star${\it Dipartimento di Fisica, Politecnico di Torino \\
C.so Duca degli Abruzzi, 24, I-10129 Torino, and\\
Istituto Nazionale di Fisica Nucleare, Sezione di Torino, \\
Italy}

\vspace{50pt}

{ABSTRACT}

\end{center}

\medskip
We report on the gauged supergravity analysis of Type IIB vacua on
$K3\times T^2/\mathbb{Z}_2$ orientifold in the presence of
$D3-D7$--branes and fluxes. We discuss supersymmetric critical
points correspond to Minkowski vacua and the related fixing of
moduli, finding agreement with previous analysis. An important
role is played by the choice of the symplectic holomorphic
sections of special geometry which enter the computation of the
scalar potential. The related period matrix ${\Scr N}$ is
explicitly given. The relation between the special geometry and
the Born--Infeld action for the brane moduli is elucidated.

\end{titlepage}

\section{Introduction}
We report on the four dimensional gauged--supergravity description
of a class of $K3\times T^2/\mathbb{Z}_2$ orientifolds
\cite{orientifold1,orientifold2} in the presence of D3--D7
space--filling branes and three--form fluxes \cite{TT,ADFL} .\par
In superstring and M--theory compactifications, which, in the
absence of fluxes have an $N$--extended local supersymmetry, the
low--energy dynamics is encoded in an effective supergravity
theory with a certain number of matter multiplets, which describe
the degrees of freedom of both bulk and brane excitations.\par In
particular in ${\mathcal N}=2$ supergravity in $D=4$ vector and
hypermultiplets are described by special and quaternionic
geometries, describing the moduli space of these theories.\par
When fluxes are turned on the effective supergravity theory
undergoes a mass deformation which is encoded in a ``gauged
supergravity'', whose general scalar potential is given in Sect.
3. This potential is completely fixed by the underlying scalar
geometry, the period matrix, the special geometry of the vector
multiplets and by the Killing vectors of the gauged isometries of
the quaternionic and special geometries.\par A correct choice of
the period matrix (explicitly given in  Appendix B) and of the
Killing isometries, would then allow to reproduce the the flux
vacua with any number ${\mathcal N}=2,1,0$ of rigid supersymmetry
in flat space \cite{tv}-\cite{Becker:2003yv}. This is indeed the
case and the agreement is found, for the particular choices of
fluxes considered, with the compactification analysis of Tripathy
and Trivedi \cite{TT}.\par These results are especially relevant
because the supergravity effective potential can be further
generalized to incorporate other perturbative or non--perturbative
results \cite{KKLT} which may further stabilize the other moduli
and lead to satisfactory inflationary cosmologies
\cite{lateKallosh}-\cite{KTW}.
\par From the point of view of the four--dimensional ${\mathcal N}=2$ effective supergravity, open string moduli, corresponding to
D7 and D3--brane positions along $T^2$, form an enlargement of the
vector multiplet moduli--space which is locally described, in
absence of open--string moduli, by \cite{ADFL}:
\begin{equation}
\label{stu}
    \left(\frac{{\rm SU} (1,1)}{{\rm U}(1)}\right)_s\times
\left(\frac{{\rm SU}(1,1)}{{\rm U}(1)}\right)_t\times
    \left(\frac{{\rm SU}(1,1)}{{\rm U}(1)}\right)_u\,,
\end{equation}
where $s,\,t,\,u$ denote the scalars of the vector multiplets
containing the $K3$--volume and the R--R four--form on $K3$, the
$T^2$--complex structure, and the IIB axion--dilaton system,
respectively\footnote{We notice that in \cite{adft} the imaginary
parts of $u$ and $t$ were chosen to be positive. This however is
inconsistent with the positivity domain of the vector kinetic
terms which requires $s,t,u$ to have negative imaginary parts.
Indeed ${\rm Im} (s)$ and ${\rm Im} (u) $ appear as coefficients
in the kinetic terms of the $D7$ and $D3$--brane vectors.}:
\begin{eqnarray}
s&=& C_{(4)} -{\rm i}\, {\rm Vol} (K_3),\,
\nonumber\\
t&=& \frac{g_{12}}{g_{22}} -{\rm i}\,\frac{\sqrt{{\rm det}
g}}{g_{22}}\,,
\nonumber\\
u&=& C_{(0)} -{\rm i}\, e^{\varphi}\,,
\end{eqnarray}
where the matrix $g$ denotes the metric on $T^2$.

When D7--branes moduli are turned on, what is known is that ${\rm
SU}(1,1)_s$ acts as an electric--magnetic duality transformation
\cite{GZ} both on the bulk and D7--brane vector field--strengths,
while the ${\rm SU} (1,1)_u$ acts as an electric--magnetic duality
transformation on the D3--vector field--strengths. Likewise the
bulk vectors transform perturbatively under ${\rm SU}(1,1)_u
\times {\rm SU}(1,1)_t$ while the D3--brane vectors do not
transform under ${\rm SU}(1,1)_s \times {\rm SU}(1,1)_t$ and the
D7--brane vectors do not transform under ${\rm SU}(1,1)_u \times
{\rm SU}(1,1)_t$.

All this is achieved starting from the following trilinear
prepotential of special geometry \cite{adft}:
\begin{equation}\label{prepot}
    {\Scr F}(s,t,u,x^k,y^r)\,=\, stu-\frac{1}{2}\,s \,x^k
    x^k-\frac{1}{2}\,u\,
    y^r y^r\,,
\end{equation}
where $x^k$ and $y^r$ are the positions of the D7 and D3--branes
along $T^2$ respectively, $k=1,\dots, n_7$, $r=1,\dots ,n_3$, and
summation over repeated indices is understood. This prepotential
is unique in order to preserve the shift--symmetries of the
$s,\,t,\,u$ bulk complex fields up to terms which only depend on
$x$ and $y$.

The above prepotential gives the correct answer if we set either
all the $x^k$ or all the $y^r$ to zero. In this case the special
geometry describes a symmetric space:
\begin{eqnarray}\label{symanifold7}
\left(\frac{{\rm SU}(1,1)}{{\rm U}(1)}\right)_s\times \frac{{\rm
SO}(2,2+n_7)}{{\rm SO}(2)\times {\rm
SO}(2+n_7)}\,,\,\,\,\,\,\,\mbox{for
$y^r=0$}\,,\\\label{symanifold3} \left(\frac{{\rm SU}(1,1)}{{\rm
U}(1)}\right)_u\times \frac{{\rm SO}(2,2+n_3)}{{\rm SO}(2)\times
{\rm SO}(2+n_3)}\,,\,\,\,\,\,\,\mbox{for $x^k=0$}\,.
\end{eqnarray}
For both $x$ and $y$ non--vanishing, the complete K\"ahler
 manifold (of complex dimension $3+n_3+n_7$) is no longer a
 symmetric space even if it still has $3+n_3+n_7$ shift
 symmetries\footnote{The prepotential in eq. (\ref{prepot}) actually
corresponds to the homogeneous not symmetric spaces called
$L(0,n_7,n_3)$ in \cite{van}. We thank A. van Proeyen for a
discussion on this point.}.

 Note that for $x^k=0$ the manifold is predicted as a truncation of
 the manifold describing the moduli--space of $T^6/\mathbb{Z}_2$
${\mathcal N}=4$ orientifold
 in the presence of D3--branes. The corresponding symplectic
 embedding was given in \cite{DFGVT}. For $y^r=0$ the
 moduli--space is predicted by the way ${\rm SU}(1,1)_s$ acts on both
 bulk and D7 vector fields. Upon compactification of Type IIB theory
on $T^2$, the D7--brane moduli are insensitive to the further
 $K3$ compactification and thus their gravity coupling must be the same as
 for vector multiplets coupled to supergravity in $D=8$.
Indeed if $2+n$ vector multiplets
 are coupled to ${\mathcal N}=2$ supergravity in $D=8$, their
non--linear $\sigma $--model
 is \cite{SS},\cite{ADF}:
\begin{equation}
\frac{{\rm SO}(2,2+n)}{{\rm SO}(2)\times {\rm SO}(2+n)}\times
\mathbb{R}^+\,.
\end{equation}
Here $\mathbb{R}^+$ denotes the volume of $T^2$ and the other part
is the second factor in (\ref{symanifold7}). Note that in
$D=8,\,{\mathcal N}=2$ the R--symmetry is ${\rm U}(1)$ which is
the ${\rm U}(1)$ part of the $D=4,\,{\mathcal N}=2$ ${\rm U}(2)$
R--symmetry. The above considerations prove eq.
(\ref{symanifold7}).\par
 Particular care is needed  \cite{AFT} when the effective supergravity is
 extended to include gauge couplings, as a result of turning on
 fluxes in the IIB compactification \cite{PS}.\par
 The reason is that the scalar potential depends explicitly on the
 symplectic embedding of the holomorphic sections of special
 geometry, while the K\"ahler potential, being symplectic
 invariant, does not. In fact, even in the analysis without open
 string moduli \cite{ADFL}, it was crucial to consider a Calabi--Visentini
 basis where the ${\rm SO}(2,2)$ linearly acting symmetry on the bulk
 fields was ${\rm SU}(1,1)_u\times {\rm SU}(1,1)_t$
 \cite{ABCDFFM},\cite{CDFV}.\par
 In the case at hand, the choice of symplectic basis is the one
 which corresponds to the Calabi--Visentini basis for $y^r=0$,
 with the ${\rm SU}(1,1)_s $ acting as an electric--magnetic duality
 transformation \cite{ADFL}, but it is not such basis for the D3--branes
 even if the $x^k=0$. Indeed, for $x^k=0$, we must reproduce the
 mixed basis used for the $T^6/\mathbb{Z}_2$ orientifold
\cite{FP},\cite{KST} in
 the presence of D3--branes found in \cite{DFGVT}.
 We note in this respect, that the choice of the symplectic
 section made in \cite{KTW} does not determine type IIB vacua
 with the 3--form fluxes turned on. It does not correspond in fact
 to the symplectic embedding discussed in \cite{ADFL},
 \cite{DFGVT} and \cite{AFT}. The problem arises already in
the absence of branes.\par
 This report is organized as follows: in
section 2 we review the gauged supergravity description of the
model and briefly discuss the moduli stabilization of different
vacua. \par In section 3 we discuss the scalar potential in the
presence of D3--D7 moduli.\par In section 4 we consider the
relation between the ${\mathcal N}=2$ special geometry
corresponding to the D3--D7 system and the Born--Infeld action,
taking into account the Chern--Simons terms describing the
couplings among bulk and brane moduli.\par The final section is
devoted to conclusions. Appendices A and B contain some relevant
formulae for the scalar potential and the period matrix ${\Scr
N}_{\varLambda\varSigma}$ of the special geometry describing the
bulk--brane coupled system of vector multiplets.

\section{${\mathcal N}=2$ and ${\mathcal N}=1$ supersymmetric cases.}

\subsection{${\mathcal N}=2$ gauged supergravity}

We consider  the gauging of ${\mathcal N}=2$ supergravity with a
special geometry given by eq. (\ref{prepot}). Let us briefly
recall the main formulae of special K\"ahler geometry. The
geometry of the manifold is encoded in the holomorphic section
$\varOmega=(X^\varLambda,\,F_\varSigma)$ which, in the {\it
special coordinate} symplectic frame, is expressed in terms of a
prepotential ${\Scr
F}(s,t,u,x^k,y^r)=F(X^\varLambda)/(X^0)^2={\Scr
F}(X^\varLambda/X^0)$, as follows:
\begin{equation}
\varOmega = ( X^\varLambda,\,F_\varLambda=\partial F/\partial
X^\varLambda)\,.\label{specialcoordinate}
\end{equation}
 In our case ${\Scr F}$ is given by eq. (\ref{prepot}). The
K\"ahler potential $K$ is given by the symplectic invariant
expression:
\begin{equation}
K = -\log \left[{\rm i}(\overline{X}^\varLambda
F_\varLambda-\overline{F}_\varLambda X^\varLambda)\right] \,.
\end{equation}
In terms of $K$ the metric has the form
$g_{i\bar{\jmath}}=\partial_i\partial_{\bar{\jmath}}K$. The
matrices $U^{\varLambda\varSigma}$ and $\overline{{\Scr
N}}_{\varLambda\varSigma}$ are respectively given by:
\begin{eqnarray}
U^{\varLambda\varSigma}&=&e^K\, {\Scr D}_i X^\varLambda {\Scr
D}_{\bar{\jmath}} \overline{X}^\varSigma\,g^{i\bar{\jmath}}=
-\frac{1}{2}\,{\rm Im}({\Scr
N})^{-1}-e^K\,\overline{X}^\varLambda X^\varSigma\,,\nonumber\\
\overline{{\Scr N}}_{\varLambda\varSigma}&=&
\hat{h}_{\varLambda|I}\circ
(\hat{f}^{-1})^I{}_\varSigma\,,\,\,\mbox{where}\,\,\,\,
\hat{f}_{I}^\varLambda = \left(\matrix{{\Scr D}_i X^\varLambda \cr
\overline{X}^\varLambda }\right)\,;\,\,\,\,\hat{h}_{\varLambda|
I}=\left(\matrix{{\Scr D}_i F_\varLambda \cr
\overline{F}_\varLambda }\right) \,.\label{N}
\end{eqnarray}
 For our choice of ${\Scr F}$, $K$ has the following form:\begin{eqnarray}
K &=& -\log [-8\,({\rm Im}(s)\,{\rm Im}(t){\rm
Im}(u)-\frac{1}{2}\,{\rm Im}(s)\,({\rm
Im}(x)^k\,)^2-\nonumber\\&&\frac{1}{2}\,{\rm Im}(u)\,({\rm
Im}(y)^r\,)^2)] \,,
\end{eqnarray}
 with ${\rm Im}(s),\,{\rm Im}(t),\,{\rm
Im}(u)<0$ at $x^k=y^r=0$. The components
$X^\varLambda,\,F_\varSigma$ of the symplectic section which
correctly describe our problem, are chosen by performing a
constant symplectic change of basis from the one in
(\ref{specialcoordinate}) given in terms of the prepotential in
eq. (\ref{prepot}). The symplectic matrix is
\begin{eqnarray}
\left(\matrix{A & -B\cr B &
A}\right)&&\nonumber\\
A&=&\frac{1}{\sqrt{2}}\,\left(\matrix{1 &0&0&0&0&0\cr 0
&0&-1&-1&0&0\cr -1 &0&0&0&0&0\cr 0 &0&1&-1&0&0\cr 0
&0&0&0&\sqrt{2}&0\cr 0 &0&0&0&0&\sqrt{2}}\right)\,,
B\,=\,\frac{1}{\sqrt{2}}\,\left(\matrix{0 &1&0&0&0&0\cr 0
&0&0&0&0&0\cr 0 &1&0&0&0&0\cr 0 &0&0&0&0&0\cr 0 &0&0&0&0&0\cr 0
&0&0&0&0&0}\right)\,.
\end{eqnarray}
The rotated symplectic sections then become
\begin{eqnarray}
X^0 &=& \frac{1}{{\sqrt{2}}}\,(1 - t\,u +
\frac{(x^k)^2}{2})\,\,\,\,,\,\,\,\,\, X^1 = -\frac{t +
u}{{\sqrt{2}}}\,,
\nonumber \\
X^2 &=&  -\frac{1}{{\sqrt{2}}}\,({1 + t\,u -
\frac{(x^k)^2}{2}})\,\,\,\,,\,\,\,\,\,X^3 = \frac{t -
u}{{\sqrt{2}}}\,,
\nonumber \\
X^k &=& x^k\,\,\,\,,\,\,\,\,\,X^r = y^r\,,
\nonumber\\
F_0 &=& \frac{s\,\left( 2 - 2\,t\,u + (x^k)^2 \right) +
u\,(y^r)^2}{2\,{\sqrt{2}}}\,\,\,\,,\,\,\,\,\, F_1 =
  \frac{-2\,s\,\left( t + u \right)  +
  (y^r)^2}{2\,{\sqrt{2}}}\nonumber\\
F_2&=&
  \frac{s\,\left( 2 + 2\,t\,u - (x^k)^2 \right)  -
  u\, (y^r)^2}{2\,{\sqrt{2}}}\,\,\,\,,\,\,\,\,\,
 F_3 = \frac{2\,s\,\left( -t + u \right)  +
 (y^r)^2}{{2\,\sqrt{2}}}\nonumber\\
F_i &=& -
  s\,x^k
 \,\,\,\,,\,\,\,\,\,
F_r = -u\,y^r\, .
\end{eqnarray}
Note that, since $\partial X^\varLambda /\partial s =0$ the new
sections do not admit a prepotential, and the no--go theorem on
partial supersymmetry breaking \cite{Cecotti} does not apply in
this case. As in \cite{ADFL}, we limit ourselves to gauge
shift--symmetries of the quaternionic manifold of the $K3$
moduli--space. Other gaugings which include the gauge group on the
brane will be considered elsewhere.\par We will also consider
particular assignments of the gauge couplings which give
$X^2=X^3=0$, i.e. $t=u=-i$. Other choices for the gauge couplings,
allowing $u\neq -i$ are possible and we shall discuss some cases
here.

\subsection{${\mathcal N}=2$ supersymmetric critical points}

In the sequel we limit our analysis to critical points in flat
space. The ${\mathcal N}=2$ critical points  demand ${\Scr
P}^x_\varLambda=0$. This equation does not depend on the special
geometry and its solution is the same as in \cite{ADFL}, i.e.
$g_2,\,g_3\neq 0$, $g_0=g_1=0$ and $e^m_a=0$ for $a=1,2$, were the
Killing vectors gauged by the fields $A^2_\mu$ and $A^3_\mu$ are
constants and their non--vanishing components are $k^u_2=g_2$
along the direction $q^u=C^{a=1}$  and $k^u_3=g_3$ along the
direction $q^u=C^{a=2}$. The 22 fields $C^m,\,C^a$, $m=1,2,3$ and
$a=1,\dots ,19$ denote the Peccei--Quinn scalars. The vanishing of
the hyperino--variation further demands:
\begin{equation}
k^u_\varLambda\, X^\varLambda = 0 \,\,\Rightarrow
\,\,\,X^2=X^3=0\,\,\,\Leftrightarrow\,\,\,t=u \,,\,\,\,
1+t^2=\frac{(x^k)^2}{2}\,.\label{n2tu}
\end{equation}
Hence for ${\mathcal N}=2$ vacua the D7 and D3--brane positions
are still moduli while the axion--dilaton and $T^2$ complex
structure are stabilised.

\subsection{${\mathcal N}=1,0$ critical points}

 The ${\mathcal N}=1$ critical points in flat space studied in
\cite{ADFL} were first obtained by setting $g_0,\,g_1\neq 0$ and
$g_2=g_3=0$, with $k^u_0=g_0$ along the direction $q^u=C^{m=1}$
and $k^u_1=g_1$ along the direction $q^u=C^{m=2}$.

\paragraph{Constant Killing spinors.}
By imposing $\delta_{\epsilon_2}\,f=0$ for the variations of the
fermionic
fields $f$  we get the following: \\
From the hyperino variations:
\begin{eqnarray}
\delta_{\epsilon_2}\,\zeta^{A
a}&=&0\,\,\,\Rightarrow\,\,\,\,e^a_m=0\,\,m=1,2\,;\,\,a=1,\dots,
19\,,\nonumber\\
\delta_{\epsilon_2}\,\zeta^{A
}&=&0\,\,\,\Rightarrow\,\,\,\mbox{vanishing of the gravitino
variation}\,.
\end{eqnarray}
The gravitino variation vanishes if:
\begin{equation}
S_{22} = -g_0\, X^0+{\rm i}\,g_1\, X^1=0\,.
\end{equation}
From the gaugino variations we obtain:
\begin{equation}
\delta_{\epsilon_2}\,(\lambda^{\bar{\imath}})_A = 0\,\,\Rightarrow
\,\,e^{\frac{K}{2}}\,{\Scr P}^x_\varLambda\,(\partial_i
X^\varLambda+(\partial_i K)\,
X^\varLambda)\,\sigma^x_{A2}\,=\,0\,,
\end{equation}
the second term (with $\partial_i K$) gives a contribution
proportional to the gravitino variation while the first term, for
$i=u,\,t,\,x^k$ respectively gives:
\begin{eqnarray}
-g_0\, \partial_u X^0+{\rm i}\,g_1\, \partial_u X^1 &=&0 \,,\nonumber\\
-g_0\, \partial_t X^0+{\rm i}\, g_1\,\partial_t X^1 &=&0 \,,\nonumber\\
-g_0\, \partial_{x^k} X^0 &=&0 \,, \nonumber\\
\end{eqnarray}
for $i=y^r$ the equation is identically satisfied. From the last
equation we get $x^k=0$ and the other two, together with
$S_{22}=0$ give $u=t=-{\rm i}$, $g_0=g_1$.

So we see that for ${\mathcal N}=1$ vacua the D7--brane
coordinates are frozen while the D3--brane coordinates remain
moduli. This agrees with the analysis of \cite{TT}. If $g_0\neq
g_1$ the above solutions give critical points with vanishing
cosmological constant but with no supersymmetry left.

More general ${\mathcal N}=1,0$ vacua can be obtained also in this
case by setting $g_2,\,g_3\neq 0$. The only extra conditions
coming from the gaugino variations for ${\mathcal N}=1$ vacua is
that $e^{a=1,2}_m=0$. This eliminates from the spectrum two extra
metric scalars $e^{a=1,2}_3$ and the $C_{a=1,2}$ axions. These
critical points preserve ${\mathcal N}=1$ or not depending on
whether $|g_0|=|g_1|$ or not.

We can describe the ${\mathcal N}=1 \to {\mathcal N}=0$ transition
with an ${\mathcal N}=1$ no--scale supergravity
\cite{noscale1,noscale2} based on a constant superpotential and a
non--linear sigma--model which is
\begin{equation}
{{\rm U} (1,1+n_3) \over {\rm U}(1)\times {\rm U} (1+ n_3)} \times
{{\rm SO} (2, 18) \over {\rm SO} (2) \times {\rm SO} (18)} \,,
\end{equation}
where the two factors come from vector multiplets and
hypermultiplets, respectively. This model has vanishing scalar
potential, reflecting the fact that there are not further scalars
becoming massive in this transition \cite{ADFL}. We further note
that any superpotential $W(y)$ for the D3 brane coordinates would
generate a potential \cite{FPorr} term
\begin{equation}
e^{K} \, K^{y\bar y}\, \partial_y W \partial_{\bar y} \bar W\,,
\end{equation}
which then would require the extra condition $\partial_y W =0$ for
a critical point with vanishing vacuum energy.

The residual moduli space of K3 metrics at fixed volume is locally
given by
\begin{equation}
{{\rm SO} (1,17) \over {\rm SO} (17)}\,.
\end{equation}
We again remark that we have considered vacua with vanishing
vacuum energy. We do not consider here the possibility of other
vacua with non--zero vacuum energy, as i.e. in \cite{KTW}.
\subsection{More general vacua}
There are more general critical points defined by values of
$t,\,u$ different form $-i$ and depending on ratios of fluxes. Let
us give an instance of this for the ${\mathcal N}=2$ preserving
vacua.\par Consider the situation with generic flux
$f^p{}_\varLambda$, $p=(m,a),\,\varLambda=0,\dots, 3$, which
corresponds to the charge--couplings:
\begin{equation}
\nabla_\mu C^p=\partial_\mu
C^p+f^p{}_\varLambda\,A_\mu^\varLambda\,.
\end{equation}
For a ${\mathcal N}=2$ vacuum, for the vanishing of the gravitino
and gaugino variations, we need ${\Scr P}^x_\varLambda=0$, where
\begin{equation}
{\Scr P}^x_\varLambda=\omega_u^x\, k^u_\varLambda\equiv
\omega_p^x\, f^p_\varLambda\,.
\end{equation}
From the hyperino variations we have
\begin{equation}
k^u_\varLambda X^\varLambda=f^p_\varLambda X^\varLambda=0\,.
\end{equation}
We take $\varLambda=2,3$ with $f^p_{2,3}\neq 0$ for $p=a$,
$(a=1\dots , 19)$ and $f^p_{\varLambda}=0$ otherwise. The hyperino
variation then is:
\begin{equation}
f^a{}_2 \, X^2+f^a{}_3 \, X^3=0\,.
\end{equation}
Setting $f^a{}_2=\alpha\, f^a{}_3$ we obtain
\begin{equation}
f^a{}_3 \, (\alpha\,X^2+ X^3)=0\,,
\end{equation}
that is
\begin{equation}
\frac{X^3}{X^2}=\frac{u-t}{1+tu-\frac{(x^k)^2}{2}}=-\alpha=-\frac{f^a{}_2}{f^a{}_3}\,.\label{tu}
\end{equation}
 The condition ${\Scr P}^x_\varLambda =0$ on the other hand
implies
\begin{equation}
e^x{}_a \, f^a{}_{2,3}=0\,,
\end{equation}
but since $f^a{}_2=\alpha\, f^a{}_3$ then the above equation  is
equivalent to the following single condition
\begin{equation}
e^x{}_a \, f^a{}_{2}=0\,,
\end{equation}
namely it fixes only one triplet of metric moduli.\par This vacuum
preserves ${\mathcal N}=2$ supersymmetry  with one massive vector
multiplet corresponding to a combination of $A^2_\mu $ and
$A^3_\mu$. Moreover condition (\ref{tu}) fixes the $T^2$ complex
structure modulus in terms of the axion--dilaton and the $x^k$
moduli of the D7 brane coordinates. Note that in the previous
solution \cite{adft} $X^2=X^3=0$, $u=t$,
$t^2=-1+\frac{(x^k)^2}{2}$ and $x^k$ were still unfixed. For
$\alpha=0$ or $\infty$ we get $X^3$ or $X^2$ vanishing which
corresponds to the example given in \cite{TT}.
\section{The potential}

The general form of the ${\mathcal N}=2$ scalar potential is:
\begin{equation}
V = 4\, e^K  h_{uv} k^u_\varLambda k^v_\varSigma
\,X^\varLambda\,\overline{X}^\varSigma+e^K g_{i\bar{\jmath}}\,
k^i_\varLambda k^{\bar{\jmath}}_\varSigma \,X^\varLambda\,
\overline{X}^\varSigma+e^K (U^{\varLambda\varSigma}-3\, e^K
\,X^\varLambda \overline{X}^\varSigma){\Scr P}^x_\varLambda\,{\Scr
P}^x_\varSigma\,,
\end{equation}
where the second term is vanishing for abelian gaugings. Here
$h_{uv}$ is the quaternionic metric and $k_\varLambda^u$ the
quaternionic Killing vector of the hypermultiplet $\sigma$--model.

The scalar potential, at the extremum of the $e^a_m$ scalars, has
the following form\footnote{Note that there is a misprint in eq.
(5.1) of ref. \cite{ADFL}. The term $e^{2\phi} \, e^{\tilde K} \,
g_0 \, g_1 ( X_0 \bar X_1 + X_1 \bar X_0 )$ is actually absent}:
\begin{eqnarray}
\label{potential} V&=& 4\,e^{2\,\phi}\, e^{K}\,
\left[\sum_{\varLambda=0}^3\, (g_\varLambda)^2\,
|X^\varLambda|^2+\frac{1}{2}\,
(g_0^2+g_1^2)(t-\bar{t})\left((u-\bar{u})-\frac{1}{2}\,\frac{(
x^k-\bar{x}^k)^2}{(t-\bar{t})}\right)\right.\nonumber\\&&\left. +
\frac{(y^r-\bar{y}^r)^2}{8\,(s-\bar{s})(u-\bar{u})}\left(g_0^2\,(\bar{u}
x^k-\bar{x}^k u)^2+g_1^2\,(x^k-\bar{x}^k)^2\right) \right]\,.
\end{eqnarray}
From the above expression we see that in the ${\mathcal N}=2$
case, namely for $g_0=g_1=0$, the potential depends on $y^r$ only
through the factor $e^{K}$ and vanishes identically in $y^r$ for
the values of the $t,u$ scalars given in (\ref{n2tu}), for which
$X^2=X^3=0$. If $g_0$ or $g_1$ are non--vanishing (${\mathcal
N}=1,\,0$ cases) the extremisation of the potential with respect
to $x^k$, namely $\partial_{x^k} V=0$ fixes $x^k=0$. For $x^k=0$
the potential depends on $y^r$ only through the factor $e^{K}$ and
vanishes identically in $y^r$ for $t=u=-{\rm i}$.
\section{Special coordinates, solvable coordinates and B.I. action}
\label{veryspecial} The prepotential for the spacial geometry of
the $D3-D7$ system, given in (\ref{prepot}), was obtained in
\cite{adft}, by using arguments based on duality symmetry, four
dimensional Chern--Simons terms coming from the p--brane couplings
as well as couplings of vector multiplets in $D=4$ and $D=8$.\par
A similar result was advocated in \cite{fms,abfpt} by performing
first a $K3$ reduction to $D=6$ and then further compactifying the
theory to $D=4$ on $T^2$.\par The subtlety of this derivation is
that the naive Born--Infeld action derived for $D5$ and $D9$
branes in $D=6$ gives kinetic terms for the scalar fields which,
at the classical level, are inconsistent with ${\mathcal N}=2$
supersymmetry. This is a consequence of the fact that anomalies
are present in the theory, as in the $D=10$ case. The mixed
anomaly local counterterms are advocated to make the Lagrangian
${\mathcal N}=2$ supersymmetric in $D=4$.\par Therefore the
corrected Lagrangian, in the original brane coordinates is highly
non--polinomial. In fact the original Born--Infeld, Chern--Simons
naive (additive) classical scalar action
\begin{eqnarray}
&&\frac{|\partial s^\prime+ c^r \partial
d^r|^2}{(s^\prime-\bar{s}^\prime)^2}+\frac{|\partial u^\prime+ a^i
\partial b^i|^2}{(u^\prime-\bar{u}^\prime)^2}+\frac{|t^\prime\,\partial d^r+\partial
c^r|^2}{(s^\prime-\bar{s}^\prime)\,(t^\prime-\bar{t}^\prime)}+\frac{|t^\prime\,\partial
b^i+\partial
a^i|^2}{(u^\prime-\bar{u}^\prime)\,(t^\prime-\bar{t}^\prime)}+\frac{|\partial
t^\prime|^2}{(t^\prime-\bar{t^\prime})^2}\,,\nonumber\\
&&\phantom{aaaaa}s^\prime=s-\frac{1}{2}\,d^r
y^r\,\,;\,\,\,u^\prime=u-\frac{1}{2}\,b^i x^i\,\,;\,\,t^\prime =t\,,\nonumber\\
&&\phantom{aaaaa}x^i=a^i+t \,b^i\,\,;\,\,\,y^r=c^r+t\,d^r\,,
\label{bia}
\end{eqnarray}
has a metric which was shown \cite{abfpt} to be K\"ahler with
K\"ahler potential\footnote{$Y_{SK}$ differs by a factor $-i$ from
the special geometry formula obtained from the prepotential in
\ref{prepot}.}
\begin{eqnarray}
K&=&
-\log\left[(s-\bar{s})(t-\bar{t})-\frac{1}{2}\,(y^r-\bar{y}^r)^2\right]-
\log\left[(u-\bar{u})(t-\bar{t})-\frac{1}{2}\,(x^i-\bar{x}^i)^2\right]+\log (t-\bar{t})\nonumber\\
&&=-\log Y_{SK}-\log(1+\frac{X_4}{Y_{SG}})\,,
\end{eqnarray}
where
\begin{eqnarray}
X_4&=&\frac{(x^i-\bar{x}^i)^2 (y^r-\bar{y}^r)^2}{4\,
(t-\bar{t})}\nonumber\\Y_{SK}&=&(s-\bar{s})(t-\bar{t})(u-\bar{u})-\frac{1}{2}\,(u-\bar{u})(y^r-\bar{y}^r)^2-
\frac{1}{2}\,(s-\bar{s})(x^i-\bar{x}^i)^2\,,\nonumber\\
\end{eqnarray}
where here and in the following summation over repeated indices is
understood.  Therefore the correction to the scalar metric in the
brane coordinates is:
\begin{eqnarray}
\partial_p\partial_{\bar{q}}\Delta K&=&
\partial_p\partial_{\bar{q}}\log(1+\frac{X_4}{Y_{SG}})\,.
\end{eqnarray}
 It is clear that the classical
brane coordinates are not good ``supersymmetric'' coordinates, in
that the corrected action is not polynomial in them. From the fact
that the combined system is a homogeneous space, we indeed expect
that suitable  coordinates exist such that the quantum corrected
(${\mathcal N}=2$ supersymmetric) action has a simple polynomial
dependence on them, including the interference term. Such
coordinates do indeed exist and allow to write the combined
Born--Infeld action and supersymmetric counterterms, in a manifest
supersymmetric way. Modulo field redefinitions, these coordinates
reduce to the standard brane coordinates when either the $D3$ or
the $D7$--branes are absent, in which cases the homogeneous space
becomes a symmetric space. This parametrization in terms of
``supersymmetric'' coordinates, corresponds to the solvable Lie
algebra description of the manifold first introduced by
Alekseevski \cite{alek}\cite{c}, which we shall discuss in what
follows. In Alekseevski's notation the manifold under
consideration is of type $K(n_3,n_7)$ which can be written as:
\begin{eqnarray}
K(n_3,n_7)&=& W(g_\alpha,\,h_\alpha,\,Y^\pm,\,Z^\pm)\,,\nonumber\\
{\rm dim}(Y^\pm)&=&n_3\,\,;\,\,\,{\rm dim}(Z^\pm) =  n_7\,,
\end{eqnarray}
where $n_3$ and $n_7$ denote the number of $D3$ and $D7$--branes
respectively. Our identification of the scalar fields with
solvable parameters is described by the following expression for a
generic solvable Lie algebra element:
\begin{eqnarray}
Solv &=&\{\sum_{\alpha=t,u,s}\varphi^\alpha h_\alpha
+\hat{\theta}_t g_t+\theta_u g_u+\theta_s g_s+y^{r\pm
}Y^{\pm}_r+z^{i\pm
}Z^{\pm}_i\}\,,\nonumber\\
&&\phantom{aaaaaa}\hat{\theta}_t = \theta_t+y^{r+}\,
y^{r-}+z^{i+}\, z^{i-}\,,
\end{eqnarray}
where $(y^{r+},y^{r-})$ and $(z^{i+},z^{i-})$ are related to the
real and imaginary parts of the $D3$ and $D7$--branes complex
coordinates along $T^2$. The non trivial commutation relations
between the above solvable generators are:
\begin{eqnarray}
[h_t,Y^\pm]&=&
\frac{1}{2}\,Y^\pm\,\,\,;\,\,\,\,\,[h_t,Z^\pm]=\frac{1}{2}\,
Z^\pm\,,\nonumber\\
\left[ h_s,Y^\pm \right] &=&\pm\frac{1}{2}\,Y^\pm\,\,\,;\,
\,\,\,\,\left[h_u,Z^\pm\right]=\pm\frac{1}{2}\,
Z^\pm\,,\nonumber\\
\left[g_s,Y^-\right]&=&Y^+\,\,\,;\,\,\,\,\,\left[g_u,Z^-\right]=
Z^+\,,\nonumber\\
\left[Y^+_r,Y^-_s\right]&=&\delta_{rs}\,g_t\,\,\,;\,\,\,\,\,\left[Z^+_i,Z^-_j\right]=
\delta_{ij}\,
g_t\,\,;\,\,\,r,s=1,\dots,n_3\,\,i,j=1,\dots,n_7\,,\nonumber\\
\left[h_\alpha,
g_\alpha\right]&=&g_\alpha\,\,;\,\,\,\alpha=t,u,s\,.
\label{comkal}
\end{eqnarray}
We exponentiate the solvable algebra using the following
coset-representative:
\begin{eqnarray}
L&=&e^{\theta_s g_s}\,e^{y^{r-} Y^-_r}\,e^{y^{r+}
Y^+_r}\,e^{\theta_u g_u}\,e^{z^{i-} Z^-_i}\,e^{z^{i+} Z^+_i}\, e^{
\hat{\theta}_t\,g_t}\,e^{\varphi^\alpha h_\alpha}\,.
\end{eqnarray}
The order of the exponentials in the coset representative and the
particular parameter $\hat{\theta}_t$ used for $g_t$, have been
chosen in such a way that the axions $\theta_s,
\,\theta_t,\,\theta_u,\,y^{r+},\,z^{i+}$ appear in the resulting
metric only covered by derivatives. The metric reads:
\begin{eqnarray}
ds^2&=&
(d\varphi_\alpha)^2+e^{-2\varphi_t}\,\left(d\theta_t+\frac{1}{2}\,d\theta_u
(z^{-})^2+\frac{1}{2}\,d\theta_s (y^{-})^2+z^{i-}\,
dz^{i+}+y^{r-}\, dy^{r+} \right)^2+\nonumber\\
&&
e^{-2\varphi_u}\,d\theta_u^2+e^{-2\varphi_s}\,d\theta_s^2+e^{-\varphi_t-\varphi_u}\,
(dz^{i+}+d\theta_u\,z^{i-})^2+e^{-\varphi_t+\varphi_u}\,
(dz^{i-})^2+\nonumber\\&& e^{-\varphi_t-\varphi_s}\,
(dy^{r+}+d\theta_s\,y^{r-})^2+e^{-\varphi_t+\varphi_s}\,
(dy^{r-})^2 \nonumber\\&&(z^{+})^2 \equiv
\sum_{i=1}^{n_7}(z^{i+})^2\,;
\,\,(y^{+})^2\equiv\sum_{r=1}^{n_3}(y^{r+})^2\,.
\end{eqnarray}
Identifying the axionic coordinates
$\theta_s,\,\theta_t,\,\theta_u,\,y^{r+},\,z^{i+}$ with the real
part of the special coordinates $s,\,t,\,u,\,y^r,\,x^i$, and
comparing the corresponding components of the metric one easily
obtains the following relations between the solvable coordinates
and the special coordinates:
\begin{eqnarray}
s&=&\theta_s-\frac{i}{2}\,e^{\varphi_s}\,\,;\,\,\,u=\theta_u-\frac{i}{2}\,
e^{\varphi_u}\,,\nonumber\\
t&=&\theta_t-\frac{i}{2}\,\left(e^{\varphi_t}+\frac{1}{2}
\,e^{\varphi_u}\,(z^{-})^2+\frac{1}{2}\,e^{\varphi_s}\,
(y^{-})^2\right)\,,\nonumber\\
x^i&=&z^{i+}+\frac{i}{2}\, e^{\varphi_u}\,
z^{i-}\,\,;\,\,\,\,y^r=y^{r+}+\frac{i}{2}\,e^{\varphi_s}\,
y^{r-}\,.
\end{eqnarray}
 Note that the classical B--I+C--S action (\ref{bia}),
with no interference term in the $D3$ ($c,\,d$) and $D7$ ($a,\,b$)
brane coordinates is still described by a homogeneous manifold
spanned by the following  $2\,n_3+2\,n_7+6$ isometries:
\begin{eqnarray}
u&\rightarrow & e^{\lambda_u}\,u\,\,;\,\,\,\delta u=u_0+a_0^i
b^i\,,\nonumber\\
s&\rightarrow & e^{\lambda_s}\,s\,\,;\,\,\,\delta s=s_0+c_0^r
d_r\,,\nonumber\\
t&\rightarrow & e^{\lambda_t}\,t\,\,;\,\,\,\delta t=t_0\,,\nonumber\\
c^r&\rightarrow &
e^{\frac{\lambda_s+\lambda_t}{2}}\,c^r\,\,;\,\,\,\delta c^r=t_0\,
d^r\,,\nonumber\\
d^r&\rightarrow &
e^{\frac{\lambda_s-\lambda_t}{2}}\,d^r\,\,;\,\,\,\delta
d^r=d_0^r\,,\nonumber\\
a^i&\rightarrow &
e^{\frac{\lambda_u+\lambda_t}{2}}\,a^i\,\,;\,\,\,\delta
a^i=a_0^i+t_0\,
b^i\,,\nonumber\\
b^i&\rightarrow &
e^{\frac{\lambda_u-\lambda_t}{2}}\,b^i\,\,;\,\,\,\delta
b^i=b_0^i\,.
\end{eqnarray}
The underlying homogeneous space is generated by the following
rank 3 solvable Lie algebra
$\{T_a^i,\,T_b^i,\,T_c^r,\,T_d^r,\,h_s,\,h_t,\,h_u,\,g_s,\,g_t,\,g_u\}$
whose non trivial commutation relations are:
\begin{eqnarray}
\left[T_a^i,\,T_b^j\right]&=&\delta^{ij}\,g_u\,\,;\,\,\,\,
\left[T_c^r,\,T_d^s\right]=\delta^{rs}\,g_s\nonumber\\
\left[T_b^i,\,g_t\right]&=&T_a^i\,\,;\,\,\,\,\left[T_d^r,\,g_t\right]=T_c^r\nonumber\\
\left[h_\alpha,\,g_\alpha\right]&=&g_\alpha\,\,\,\,\,\alpha=s,t,u\nonumber\\
\left[h_s,\,T^r_d\right]&=&\frac{1}{2}\,T^r_d\,\,;\,\,\,\,
\left[h_s,\,T^r_c\right]=\frac{1}{2}\,T^r_c\nonumber\\
\left[h_u,\,T^i_b\right]&=&\frac{1}{2}\,T^i_b\,\,;\,\,\,\,
\left[h_u,\,T^i_a\right]=\frac{1}{2}\,T^i_a\nonumber\\
\left[h_t,\,T^r_d\right]&=&-\frac{1}{2}\,T^r_d\,\,;\,\,\,\,
\left[h_t,\,T^r_c\right]=\frac{1}{2}\,T^r_c\nonumber\\
\left[h_t,\,T^i_b\right]&=&-\frac{1}{2}\,T^i_b\,\,;\,\,\,\,
\left[h_t,\,T^i_a\right]=\frac{1}{2}\,T^i_a\,,
\end{eqnarray}
where the nilpotent generators have been labelled by the
corresponding axionic scalar fields. This space is not a subspace
of the original quanternionic space, but it becomes so if we set
either $a,b=0$ and exchange the role of $s$ and $t$ or if we set
$c,d=0$ and exchange the role of $u$ and $t$.\par The amazing
story is that the coordinates in $D=4$ corresponding to the
supersymmetric theory, deform this space into an other homogeneous
space generated by the isometries in (\ref{comkal}) which
corresponds to an ${\mathcal N}=2$ special geometry.\par The
relation between the solvable Lie algebra generators
$\{T_a^i,\,T_b^i,\,T_c^r,\,T_d^r,\,h_s,\,h_t,\,h_u$
$,\,g_s,\,g_t,\,g_u\}$ corresponding to the classical coordinates
and the solvable generators \\ $\{Y^\pm,\,Z^\pm,\,
h_\alpha,\,g_\alpha\}$ corresponding to the ``supersymmetric''
coordinates is the following:
\begin{eqnarray}
T_a^i&=&\hat{Z}^{i+}\,\,;\,\,\,T_b^i=\hat{Z}^{i-}\,,\nonumber\\
T_c^r&=&\hat{Y}^{r+}\,\,;\,\,\,T_d^r=\hat{Y}^{r-}\,,
\end{eqnarray}
where $\hat{Y}$ and $\hat{Z}$ are the generators with opposite
grading with respect to $Y$ and $Z$ respectively. It can be shown
that in the manifold $K(n_3,n_7)$,  $\hat{Y}$ or $\hat{Z}$ are
isometries only if $n_7=0$ or $n_3=0$ respectively. Indeed in
these two cases the manifold is symmetric and each solvable
nilpotent isometry has a ``hidden'' counterpart with opposite
grading. Otherwise the manifold spanned by the classical
coordinates and the manifold parametrized by the
``supersymmetric'' ones are in general different.
\section{Conclusions}
The present investigation allows us to study in a fairly general
way the potential for the 3--form flux compactification, in
presence of both bulk and open string moduli. In absence of fluxes
the D3, D7 dependence of the K\"ahler potential is rather
different since this moduli couple in different ways to the bulk
moduli.

Moreover, in the presence of 3--form fluxes which break ${\mathcal
N}=2\rightarrow {\mathcal N}=1,0$ the D7 moduli are stabilised
while the D3 moduli are not. For small values of the coordinates
$x^k$, $y^r$ the dependence of their kinetic term is (for
$u=t=-{\rm i}$), $-(\partial_\mu \bar{y}^r\partial^\mu y^r)/{\rm
Im}(s)$ for the D3--brane moduli, and $-(\partial_\mu
\bar{x}^k\partial^\mu x^k)$ for the D7--brane moduli. This is in
accordance with the suggestion of \cite{lateKallosh}. Note that
the above formulae, at $x=0,\,u=t=-{\rm i}$ are true up to
corrections $O(\frac{{Im}(y)^2}{{Im}(s)})$, since $y$ and $s$ are
moduli even in presence of fluxes. The actual dependence of these
terms on the compactification volume is important in order to
further consider models for inflatons where the terms in the
scalar potential allow to stabilise the remaining moduli.

Finally, we have not considered here the gauging of compact gauge
groups which exist on the brane world--volumes. This is, for
instance, required \cite{dterm1,dterm2,KTW} in models with hybrid
inflation \cite{hybrid}. This issue will be considered elsewhere.

\vskip 1.5truecm

\noindent{\bf Acknowledgements. } Work supported in part by the
European Community's Human Potential Program under contract
HPRN-CT-2000-00131 Quantum Space-Time, in which  R. D'A. is
associated to Torino University. The work of S.F. has been
supported in part by European Community's Human Potential Program
under contract HPRN-CT-2000-00131 Quantum Space-Time, in
association with INFN Frascati National Laboratories and by D.O.E.
grant DE-FG03-91ER40662, Task C.

\appendix
\section*{Appendix A \ \  Some relevant formulae. }\label{appendiceA}
\setcounter {equation}{0} \addtocounter{section}{1}

We are interested in gauging the 22 translations in the coset
${\rm SO}(4,20)/({\rm SO}(3,19)\times {\rm O}(1,1))$. Let us
denote by $L$ the coset representative of ${\rm SO}(3,19)/{\rm
SO}(3)\times {\rm SO}(19)$. It will be written in the form:
\begin{equation}
L = \left(\matrix{(1+{\bf e}\,{\bf e}^T)^{\frac{1}{2}} & -{\bf
e}\cr -{\bf e}^T & (1+{\bf e}^T{\bf e})^{\frac{1}{2}} }\right)\,,
\end{equation}
where ${\bf e}=\{e^m{}_a\}$, ${\bf e}^T=\{e^a{}_m\}$ , $m=1,2,3$
and $a=1,\dots , 19$, are the coordinates of the manifold. The
$22$ nilpotent Peccei--Quinn generators are denoted by
$\{Z_m,\,Z_a\}$ and the gauge generators are:
\begin{equation}
t_\varLambda = f^m{}_\varLambda Z_m+ h^a{}_\varLambda Z_a\,,
\end{equation}
the corresponding Killing vectors have non vanishing components:
$k^m_\varLambda=f^m{}_\varLambda$ and
$k^a_\varLambda=h^a{}_\varLambda$. The moment maps are:
\begin{equation}
{\Scr P}^x_\varLambda =
\sqrt{2}\,\left(e^{\phi}\,(L^{-1})^x{}_m\,f^m{}_\varLambda
+e^{\phi}\,(L^{-1})^x{}_a\,h^a{}_\varLambda\right)\,,
\end{equation}
where $\phi$ is the $T^2$ volume modulus \cite{ADFL}:
$e^{-2\,\phi}={\rm Vol}(T^2)$ and $x=1,2,3$. The metric along the
Peccei--Quinn directions $I=(m,a)$ is:
\begin{equation}
h_{IJ}  = e^{2\,\phi}\,(\delta_{IJ}+2\,e^a{}_I e^a{}_J)\,.
\end{equation}
The potential has the following form:
\begin{eqnarray}
V&=&4\,
e^{2\,\phi}\,\left(f^m{}_\varLambda\,f^m{}_\varSigma+2\,e^a{}_m
e^a{}_n\,f^m{}_\varLambda\,f^n{}_\varSigma+h^a{}_\varLambda\,
h^a{}_\varSigma\right)\,\bar{L}^\varLambda\,
L^\varSigma \nonumber\\
&& +
2\,e^{2\,\phi}\,\left(U^{\varLambda\varSigma}-3\,\bar{L}^\varLambda\,
L^\varSigma\right)\,\left(f^m{}_\varLambda\,f^m{}_\varSigma+e^a{}_m
e^a{}_n\,f^m{}_\varLambda\,f^n{}_\varSigma \right.
\nonumber\\
&&\left. + 2\,[(1+{\bf e}\,{\bf e}^T)^{\frac{1}{2}}]^n{}_m
e^n{}_a\, f^m{}_{(\varLambda}\,h^a{}_{\varSigma)}+e^n{}_a
e^n{}_b\,h^a{}_\varLambda\,h^b{}_\varSigma\right) \,. \label{pot}
\end{eqnarray}
In all the models we consider, at the extremum point of the
potential in the special K\"ahler manifold the following condition
holds: $\left(U^{\varLambda\varSigma}-3\,\bar{L}^\varLambda\,
L^\varSigma\right)_{|0}\,f^m{}_{(\varLambda}\,h^a{}_{\varSigma)}=0$.
As a consequence of this, as it is clear from (\ref{pot}), the
potential in this point depends on the metric scalars $e^m_a$ only
through quadratic terms in the combinations
$e^m{}_a\,h^a{}_\varLambda$ and $ e^a{}_m\,f^m{}_\varLambda$.
Therefore $V$ is extremised with respect to the $e^m_a$ scalars
once we restrict ourselves to the moduli defined as follows:
\begin{equation}
\mbox{moduli:} \qquad  e^m{}_a\,h^a{}_\varLambda
=e^a{}_m\,f^m{}_\varLambda =0 \,.
\end{equation}
The vanishing of the potential implies
\begin{equation}
\left(U^{\varLambda\varSigma}-\,\bar{L}^\varLambda\,
L^\varSigma\right)_{|0}\,f^m{}_{(\varLambda}\,f^m{}_{\varSigma)}
+2 \left(\bar{L}^\varLambda\,
L^\varSigma\right)_{|0}\,h^a{}_{(\varLambda}\,h^a{}_{\varSigma)}=0
\,.
\end{equation}
Furthermore, one may notice that, as in \cite{ADFL}, the following
relations hold in all the models under consideration:
\begin{equation}
\left(U^{\varLambda\varSigma}-\,\bar{L}^\varLambda\,
L^\varSigma\right)_{|0}\,f^m{}_{(\varLambda}\,f^m{}_{\varSigma)}
=\left(\bar{L}^\varLambda\,
L^\varSigma\right)_{|0}\,h^a{}_{(\varLambda}\,h^a{}_{\varSigma)}=0
\,.
\end{equation}
Our analysis is limited to the case in which the only
non--vanishing $f$ and $h$ constants are:
\begin{eqnarray}
f^1{}_0 &=& g_0 \,\,;\,\,\,f^2{}_1=g_1 \,\,;\,\,\,h^1{}_2=g_2
\,\,;\,\,\,h^2{}_3=g_3 \,\,;\,\,\,h^{2+k}{}_{3+k}=g^k_4
\nonumber\\h^{2+n_7+r}{}_{3+n_7+r}&=&g^r_5 \,.
\end{eqnarray}
\appendix
\section*{Appendix B \ \  The matrix ${\Scr N}$. }\label{appendiceB}
\setcounter {equation}{0} \addtocounter{section}{1} Using the
special geometry formula (\ref{N}) it is possible to compute the
matrix ${\Scr  N}_{\varLambda\varSigma}$ for any choice of the
symplectic section, including those cases for which no
prepotential exists. For the sake of simplicity we will suppress
the indices $k$ and $r$ in $x^k$ and $y^r$ by considering the case
$n_3=n_7=1$.
 Moreover
we will express the complex coordinates in terms of their real and
imaginary parts:
\begin{equation}
s= s_1+{\rm i}\, s_2\,\,;\,\,\,t= t_1+{\rm i}\, t_2\,\,;\,\,\,u=
u_1+{\rm i}\, u_2 \,\,;\,\,\,x= x_1+{\rm i}\, x_2\,\,;\,\,\,y=
y_1+{\rm i}\, y_2
\end{equation}
Let the $D7$ and $D3$ brane vectors correspond to the values
$\varLambda=4,5$ respectively. We list below the independent
components of the real and imaginary parts of ${\Scr  N}$:
{\small\begin{flushleft}\begin{eqnarray} {\rm Re}({\Scr
N})_{0,0}&=& {s_1} -
 \frac{1}{2}\,{u_1}\,{ {y_1}}^2 +
    \frac{ {u_2}\,\left( -2 + 2\, {t_1}\, {u_1} -
         { {x_1}}^2 \right) \, {y_1}\, {y_2}}{2\, {t_2}\,
         {u_2} - { {x_2}}^2} -\nonumber\\&&
    \frac{\left( -1 +  {t_1}\, {u_1} - \frac{1}{2}\,{ {x_1}}^2 \right) \,
       \left( 2\, {t_1}\,{ {u_2}}^2 +
          {x_2}\,\left( -2\, {u_2}\, {x_1} +
             {u_1}\, {x_2} \right)  \right) \,{ {y_2}}^2}{{\left(
          -2\, {t_2}\, {u_2} + { {x_2}}^2 \right) }^2} \nonumber\\
{\rm Re}({\Scr  N})_{0,1}&=&\frac{ {y_1}\,\left( -2\, {t_2}\,
{u_2}\, {y_1} +
      { {x_2}}^2\, {y_1} +
      4\,\left(  {t_1} +  {u_1} \right) \, {u_2}\, {y_2}
      \right) }{8\, {t_2}\, {u_2} -
      4\,{ {x_2}}^2}+\nonumber\\&&
      \frac{{y_2}^2\,\left( 2\,{ {u_2}}^2\,
       \left( 2 - 2\, {t_1}\,
          \left(  {t_1} + 2\, {u_1} \right)  + { {x_1}}^2 \right)
          + 4\,\left(  {t_1} +  {u_1} \right) \, {u_2}\,
        {x_1}\, {x_2}\right)}
    {4\,{\left( -2\, {t_2}\, {u_2} + { {x_2}}^2 \right) }^2}
    +\nonumber\\&&\frac{
      \left( 2 - 2\, {u_1}\,\left( 2\, {t_1} +  {u_1} \right)  +
         { {x_1}}^2 \right) \,{ {x_2}}^2  \,{ {y_2}}^2}
    {4\,{\left( -2\, {t_2}\, {u_2} + { {x_2}}^2 \right)
    }^2}\nonumber\\
    {\rm Re}({\Scr  N})_{0,2}&=&\frac{1}{2}\, {u_1}\,{ {y_1}}^2 +
    \frac{ {u_2}\,\left( 2\, {t_1}\, {u_1} -
         { {x_1}}^2 \right) \, {y_1}\, {y_2}}{-2\,
         {t_2}\, {u_2} + { {x_2}}^2}
        +\nonumber\\&&
    \frac{\left(  {t_1}\, {u_1} -\frac{1}{2}\, { {x_1}}^2 \right) \,
       \left( 2\, {t_1}\,{ {u_2}}^2 +
          {x_2}\,\left( -2\, {u_2}\, {x_1} +
             {u_1}\, {x_2} \right)  \right) \,{ {y_2}}^2}{{\left(
          -2\, {t_2}\, {u_2} + { {x_2}}^2 \right) }^2}\nonumber\\
    {\rm Re}({\Scr  N})_{0,3}&=& \frac{ {y_1}\,\left( -2\, {t_2}\, {u_2}\, {y_1} +
      { {x_2}}^2\, {y_1} +
      4\,\left(  {t_1} -  {u_1} \right) \, {u_2}\, {y_2}
      \right) }{8\, {t_2}\, {u_2} -
      4\,{ {x_2}}^2}+\nonumber\\&&
\frac{-\left(  {u_2}\,\left(  {u_2}\,
         \left( 2 + 2\,{ {t_1}}^2 - 4\, {t_1}\, {u_1} +
           { {x_1}}^2 \right)  +
        2\,\left( - {t_1} +  {u_1} \right) \, {x_1}\,
          {x_2} \right) \,{ {y_2}}^2 \right) }{2\,
    {\left( -2\, {t_2}\, {u_2} + { {x_2}}^2 \right)
    }^2}+\nonumber\\
    &&\frac{\left( 2 + 2\, {u_1}\,
       \left( -2\, {t_1} +  {u_1} \right)  + { {x_1}}^2 \right) \,
     { {x_2}}^2\,{ {y_2}}^2}{4\,
    {\left( -2\, {t_2}\, {u_2} + { {x_2}}^2 \right)
    }^2}\nonumber\end{eqnarray}\begin{eqnarray}
      {\rm Re}({\Scr  N})_{0,4}&=&- \frac{ {x_1}\, {y_2}\,
      \left( -4\, {t_2}\,{ {u_2}}^2\, {y_1} +
        2\, {t_1}\,{ {u_2}}^2\, {y_2} +
         {u_1}\,{ {x_2}}^2\, {y_2} \right) }{{\sqrt{2}}\,
      {\left( -2\, {t_2}\, {u_2} + { {x_2}}^2 \right) }^2}
   +\nonumber \\&&
\frac{ {u_2}\, {x_2}\, {y_2}\,
    \left( -2\, {x_1}\, {x_2}\, {y_1} +
      \left( 2 - 2\, {t_1}\, {u_1} + 3\,{ {x_1}}^2 \right) \,
        {y_2} \right) }{{\sqrt{2}}\,
    {\left( -2\, {t_2}\, {u_2} + { {x_2}}^2 \right) }^2}\nonumber\\
      {\rm Re}({\Scr  N})_{0,5}&=&\frac{2\, {t_2}\, {u_1}\, {u_2}\, {y_1} -
     {u_1}\,{ {x_2}}^2\, {y_1} +
     {u_2}\,\left( 2 - 2\, {t_1}\, {u_1} +
       { {x_1}}^2 \right) \, {y_2}}{{\sqrt{2}}\,
    \left( 2\, {t_2}\, {u_2} - { {x_2}}^2 \right) }\nonumber\\
      {\rm Re}({\Scr  N})_{1,1}&=& {s_1} - \frac{\left(  {t_1} +  {u_1} \right) \,
     \left( 2\,{ {u_2}}^2 + { {x_2}}^2 \right) \,{ {y_2}}^2}
     {{\left( -2\, {t_2}\, {u_2} + { {x_2}}^2 \right) }^2}\nonumber\\
      {\rm Re}({\Scr  N})_{1,2}&=&\frac{ {y_1}\,\left( 2\, {t_2}\, {u_2}\, {y_1} -
      { {x_2}}^2\, {y_1} -
      4\,\left(  {t_1} +  {u_1} \right) \, {u_2}\, {y_2}
      \right) }{8\, {t_2}\, {u_2} -
      4\,{ {x_2}}^2}+\nonumber\\&&\frac{ {u_2}\,\left(  {u_2}\,
       \left( 2 + 2\,{ {t_1}}^2 + 4\, {t_1}\, {u_1} -
         { {x_1}}^2 \right)  -
      2\,\left(  {t_1} +  {u_1} \right) \, {x_1}\, {x_2}
      \right) \,{ {y_2}}^2}{2\,
    {\left( -2\, {t_2}\, {u_2} + { {x_2}}^2 \right)
    }^2}+\nonumber\\&&\frac{\left( 2 + 2\, {u_1}\,\left( 2\, {t_1} +  {u_1} \right)  -
      { {x_1}}^2 \right) \,{ {x_2}}^2\,{ {y_2}}^2}{4\,
    {\left( -2\, {t_2}\, {u_2} + { {x_2}}^2 \right)
    }^2}\nonumber\\
      {\rm Re}({\Scr  N})_{1,3}&=&\frac{\left( 2\, {u_1}\,{ {u_2}}^2 -
       {t_1}\,{ {x_2}}^2 \right) \,{ {y_2}}^2}{{\left( -2\,
         {t_2}\, {u_2} + { {x_2}}^2 \right) }^2}\nonumber\\
      {\rm Re}({\Scr  N})_{1,4}&=&- \frac{\left( 2\,{ {u_2}}^2\, {x_1} +
        2\,\left(  {t_1} +  {u_1} \right) \, {u_2}\,
          {x_2} +  {x_1}\,{ {x_2}}^2 \right) \,{ {y_2}}^2
      }{{\sqrt{2}}\,{\left( -2\, {t_2}\, {u_2} + { {x_2}}^2
          \right) }^2}\nonumber\\
      {\rm Re}({\Scr  N})_{1,5}&=&\frac{{\sqrt{2}}\,\left( 2\, {t_2}\, {u_2}\, {y_1} -
      { {x_2}}^2\, {y_1} -
      2\,\left(  {t_1} +  {u_1} \right) \, {u_2}\, {y_2}
      \right) }{4\, {t_2}\, {u_2} - 2\,{ {x_2}}^2}\nonumber\\
      {\rm Re}({\Scr  N})_{2,2}&=&- {s_1} -\frac{1}{2}\,  {u_1}\,{ {y_1}}^2 +
    \frac{ {u_2}\,\left( 2 + 2\, {t_1}\, {u_1} -
         { {x_1}}^2 \right) \, {y_1}\, {y_2}}{2\, {t_2}\,
         {u_2} - { {x_2}}^2} -\nonumber\\&&
    \frac{\left( 1 +  {t_1}\, {u_1} - \frac{1}{2}\,{ {x_1}}^2 \right) \,
       \left( 2\, {t_1}\,{ {u_2}}^2 +
          {x_2}\,\left( -2\, {u_2}\, {x_1} +
             {u_1}\, {x_2} \right)  \right) \,{ {y_2}}^2}{{\left(
          -2\, {t_2}\, {u_2} + { {x_2}}^2 \right)
          }^2}\nonumber\\
      {\rm Re}({\Scr  N})_{2,3}&=&\frac{ {y_1}\,\left( 2\, {t_2}\, {u_2}\, {y_1} -
      { {x_2}}^2\, {y_1} +
      4\,\left( - {t_1} +  {u_1} \right) \, {u_2}\, {y_2}
      \right) }{8\, {t_2}\, {u_2} -
      4\,{ {x_2}}^2}+\nonumber\\&&\frac{ {u_2}\,\left(  {u_2}\,
       \left( -2 + 2\,{ {t_1}}^2 - 4\, {t_1}\, {u_1} +
         { {x_1}}^2 \right)  +
      2\,\left( - {t_1} +  {u_1} \right) \, {x_1}\, {x_2}
      \right) \,{ {y_2}}^2}{2\,
    {\left( -2\, {t_2}\, {u_2} + { {x_2}}^2 \right)
    }^2}+\nonumber\\&&\frac{\left( 2 + 4\, {t_1}\, {u_1} - 2\,{ {u_1}}^2 -
      { {x_1}}^2 \right) \,{ {x_2}}^2\,{ {y_2}}^2}{4\,
    {\left( -2\, {t_2}\, {u_2} + { {x_2}}^2 \right) }^2}\nonumber\end{eqnarray}\begin{eqnarray}
 {\rm Re}({\Scr  N})_{2,4}&=&\frac{ {x_1}\, {y_2}\,
    \left( -4\, {t_2}\,{ {u_2}}^2\, {y_1} +
      2\, {t_1}\,{ {u_2}}^2\, {y_2} +
       {u_1}\,{ {x_2}}^2\, {y_2} \right) }{{\sqrt{2}}\,
    {\left( -2\, {t_2}\, {u_2} + { {x_2}}^2 \right)
    }^2}+\nonumber\\&&\frac{ {u_2}\, {x_2}\, {y_2}\,
    \left( 2\, {x_1}\, {x_2}\, {y_1} -
      3\,{ {x_1}}^2\, {y_2} +
      2\,\left(  {y_2} +  {t_1}\, {u_1}\, {y_2} \right)
      \right) }{{\sqrt{2}}\,{\left( -2\, {t_2}\, {u_2} +
        { {x_2}}^2 \right) }^2}\nonumber\\
 {\rm Re}({\Scr  N})_{2,5}&=&\frac{-2\, {t_2}\, {u_1}\, {u_2}\, {y_1} +
     {u_1}\,{ {x_2}}^2\, {y_1} +
     {u_2}\,\left( 2 + 2\, {t_1}\, {u_1} -
       { {x_1}}^2 \right) \, {y_2}}{{\sqrt{2}}\,
    \left( 2\, {t_2}\, {u_2} - { {x_2}}^2 \right) }\nonumber\\
 {\rm Re}({\Scr  N})_{3,3}&=&- {s_1} + \frac{\left(  {t_1} -  {u_1} \right) \,
     \left( 2\,{ {u_2}}^2 - { {x_2}}^2 \right) \,{ {y_2}}^2}
     {{\left( -2\, {t_2}\, {u_2} + { {x_2}}^2 \right)
     }^2}\nonumber\\{\rm Re}({\Scr  N})_{3,4}&=&\frac{\left( 2\,{ {u_2}}^2\, {x_1} +
      2\,\left( - {t_1} +  {u_1} \right) \, {u_2}\,
        {x_2} -  {x_1}\,{ {x_2}}^2 \right) \,{ {y_2}}^2}
    {{\sqrt{2}}\,{\left( -2\, {t_2}\, {u_2} + { {x_2}}^2 \right)
        }^2}\nonumber\\ {\rm Re}({\Scr  N})_{3,5}&=&\frac{{\sqrt{2}}\,\left( 2\, {t_2}\, {u_2}\, {y_1} -
      { {x_2}}^2\, {y_1} +
      2\,\left( - {t_1} +  {u_1} \right) \, {u_2}\, {y_2}
      \right) }{4\, {t_2}\, {u_2} -
      2\,{ {x_2}}^2}\nonumber\\{\rm Re}({\Scr  N})_{4,4}&=&- {s_1} - \frac{4\, {u_2}\, {x_1}\, {x_2}\,
     { {y_2}}^2}{{\left( -2\, {t_2}\, {u_2} +
        { {x_2}}^2 \right) }^2}\nonumber\\{\rm Re}({\Scr  N})_{4,5}&=&\frac{2\, {u_2}\, {x_1}\, {y_2}}
  {-2\, {t_2}\, {u_2} + { {x_2}}^2}\nonumber\\
  {\rm Re}({\Scr  N})_{5,5}&=&-u_1
\end{eqnarray}\end{flushleft}}
As far as ${\rm Im}({\Scr  N})$ is concerned, its independent
entries are: {\scriptsize
\begin{eqnarray}
 {\rm Im}({\Scr  N})_{0,0}&=&
\frac{ {s_2}\,\left( 4 + 4\,
       \left( { {t_1}}^2 + { {t_2}}^2 \right) \,
       \left( { {u_1}}^2 + { {u_2}}^2 \right)  + { {x_1}}^4 +
      { {x_2}}^4 + 2\,{ {x_1}}^2\,\left( 2 + { {x_2}}^2 \right)
      \right) }{8\, {t_2}\, {u_2} -
      4\,{ {x_2}}^2}+\nonumber\\&&\frac{ {s_2}\, {t_2}\,
    \left( -2\, {u_1}\, {x_1}\, {x_2} +
       {u_2}\,\left(  {x_1} -  {x_2} \right) \,
       \left(  {x_1} +  {x_2} \right)  \right) }{2\, {t_2}\,
      {u_2} - { {x_2}}^2}+\nonumber\\&&- \frac{ {s_2}\, {t_1}\,
      \left( 2\, {u_2}\, {x_1}\, {x_2} +
         {u_1}\,\left( 2 + { {x_1}}^2 - { {x_2}}^2 \right)
        \right) }{2\, {t_2}\, {u_2} - { {x_2}}^2}
        +\nonumber\\&&
\frac{ {y_1}}{2}\,\left(  {u_2}\, {y_1} +
      \frac{2\,\left( 2\, {t_1}\,{ {u_2}}^2 +
            {x_2}\,\left( -2\, {u_2}\, {x_1} +
               {u_1}\, {x_2} \right)  \right) \, {y_2}}{-2\,
           {t_2}\, {u_2} + { {x_2}}^2} \right)
          +\nonumber\\&&
\frac{- {u_2}\,\left( 4\,{ {t_1}}^2\,
         \left( { {u_1}}^2 - { {u_2}}^2 \right)  +
        4\,{ {t_2}}^2\,\left( { {u_1}}^2 + { {u_2}}^2 \right)  -
        4\, {t_1}\, {u_1}\,\left( 2 + { {x_1}}^2 \right)  +
        {\left( 2 + { {x_1}}^2 \right) }^2 \right) \,{ {y_2}}^2  }
    {4\,{\left( -2\, {t_2}\, {u_2} + { {x_2}}^2 \right)
    }^2}+\nonumber\\&&
    \frac{ {x_2}\,\left(  {t_2}\,{ {u_1}}^2\, {x_2} +
       {u_2}\,\left( -2\, {t_1}\, {u_2}\, {x_1} +
         \left(  {t_1}\, {u_1} +  {t_2}\, {u_2} +
            { {x_1}}^2 \right) \, {x_2} \right)  \right) \,
    { {y_2}}^2}{{\left( -2\, {t_2}\, {u_2} +
       { {x_2}}^2 \right) }^2}+\nonumber\\&&
\frac{- { {x_2}}^3\,\left( 4\, {u_1}\, {x_1} +
         {u_2}\, {x_2} \right) \,{ {y_2}}^2  }{4\,
    {\left( -2\, {t_2}\, {u_2} + { {x_2}}^2 \right) }^2}\nonumber\\
{\rm Im}({\Scr  N})_{0,1}&=& \frac{ {s_2}\,\left( -2\,\left(
 {t_2} +  {u_2} \right) \,
        {x_1}\, {x_2} +
       {u_1}\,\left( -2 + 2\,{ {t_1}}^2 + 2\,{ {t_2}}^2 -
         { {x_1}}^2 + { {x_2}}^2 \right)  \right) }{4\, {t_2}\,
      {u_2} - 2\,{ {x_2}}^2}+\nonumber\\&&
\frac{ {s_2}\, {t_1}\,
    \left( -2 + 2\,{ {u_1}}^2 + 2\,{ {u_2}}^2 - { {x_1}}^2 +
      { {x_2}}^2 \right) }{4\, {t_2}\, {u_2} -
    2\,{ {x_2}}^2}+\nonumber\\&&
- \frac{ {y_2}\,\left( { {t_2}}^2\, {u_1}\, {u_2}\,
          {y_2} -  {t_1}\,{ {u_2}}^3\, {y_2} +
         {t_2}\,\left( 2\,{ {u_2}}^3\, {y_1} +
            {u_2}\,{ {x_2}}^2\, {y_1} -
            {u_1}\,{ {x_2}}^2\, {y_2} \right)  \right) }{{\left(
         -2\, {t_2}\, {u_2} + { {x_2}}^2 \right) }^2}
         +\nonumber\\&&
\frac{-\left( \left(  {t_1} +  {u_1} \right) \, {u_2}\,
      \left( -2 + 2\, {t_1}\, {u_1} - { {x_1}}^2 -
        { {x_2}}^2 \right) \,{ {y_2}}^2 \right) }{2\,
    {\left( -2\, {t_2}\, {u_2} + { {x_2}}^2 \right)
    }^2}+\nonumber\\&&\frac{ {x_2}\,\left( 2\,{ {u_2}}^2 + { {x_2}}^2 \right) \,
     {y_2}\,\left(  {x_2}\, {y_1} -
       {x_1}\, {y_2} \right) }{2\,
    {\left( -2\, {t_2}\, {u_2} + { {x_2}}^2 \right) }^2}
\nonumber\\
    {\rm Im}({\Scr  N})_{0,2}&=&- \frac{ {s_2}\,\left( -4 +
        4\,{ {t_1}}^2\,\left( { {u_1}}^2 + { {u_2}}^2 \right)  +
        4\,{ {t_2}}^2\,\left( { {u_1}}^2 + { {u_2}}^2 \right)  +
        { {x_1}}^4 + 2\,{ {x_1}}^2\,{ {x_2}}^2 +
        { {x_2}}^4 \right) }{8\, {t_2}\, {u_2} -
      4\,{ {x_2}}^2}+\nonumber\\&&
\frac{ {s_2}\,\left(  {t_1}\,
       \left( 2\, {u_2}\, {x_1}\, {x_2} +
          {u_1}\,\left( { {x_1}}^2 - { {x_2}}^2 \right)  \right)
       +  {t_2}\,\left( 2\, {u_1}\, {x_1}\, {x_2} +
          {u_2}\,\left( -{ {x_1}}^2 + { {x_2}}^2 \right)  \right)
          \right) }{2\, {t_2}\, {u_2} -
          { {x_2}}^2}+\nonumber\\&&
\frac{ {y_1}\,\left( -2\, {t_2}\,{ {u_2}}^2\, {y_1} +
      4\, {t_1}\,{ {u_2}}^2\, {y_2} +
      2\, {u_1}\,{ {x_2}}^2\, {y_2} +
       {u_2}\, {x_2}\,
       \left(  {x_2}\, {y_1} - 4\, {x_1}\, {y_2} \right)
      \right) }{4\, {t_2}\, {u_2} -
      2\,{ {x_2}}^2}+\nonumber\\&&
\frac{-\left(  {u_2}\,\left( 4 -
        4\,\left( { {t_1}}^2 + { {t_2}}^2 \right) \,{ {u_1}}^2 +
        4\,\left(  {t_1} -  {t_2} \right) \,
         \left(  {t_1} +  {t_2} \right) \,{ {u_2}}^2 +
        4\, {t_1}\, {u_1}\,{ {x_1}}^2 - { {x_1}}^4
        \right) \,{ {y_2}}^2 \right) }{4\,
    {\left( -2\, {t_2}\, {u_2} + { {x_2}}^2 \right)
    }^2}+\nonumber\\&&
- \frac{ {x_2}\,\left( \left(  {t_2}\,
            \left( { {u_1}}^2 + { {u_2}}^2 \right)  +
            {u_2}\,{ {x_1}}^2 \right) \, {x_2} +
         {t_1}\, {u_2}\,
         \left( -2\, {u_2}\, {x_1} +  {u_1}\, {x_2}
           \right)  \right) \,{ {y_2}}^2}{{\left( -2\, {t_2}\,
           {u_2} + { {x_2}}^2 \right) }^2}
         +\nonumber\\&&
\frac{{ {x_2}}^3\,\left( 4\, {u_1}\, {x_1} +
       {u_2}\, {x_2} \right) \,{ {y_2}}^2}{4\,
    {\left( -2\, {t_2}\, {u_2} + { {x_2}}^2 \right) }^2}
    \nonumber\\
    {\rm Im}({\Scr  N})_{0,3}&=&
\frac{ {s_2}\,\left( 2\,\left( - {t_2} +  {u_2} \right) \,
        {x_1}\, {x_2} +
       {u_1}\,\left( 2 + 2\,{ {t_1}}^2 + 2\,{ {t_2}}^2 +
         { {x_1}}^2 - { {x_2}}^2 \right)  \right) }{4\, {t_2}\,
      {u_2} - 2\,{ {x_2}}^2}+\nonumber\\&&
\frac{-\left(  {s_2}\, {t_1}\,
       \left( 2 + 2\,{ {u_1}}^2 + 2\,{ {u_2}}^2 + { {x_1}}^2 -
         { {x_2}}^2 \right)  \right)  +
    \left( 2\,{ {u_2}}^2 - { {x_2}}^2 \right) \, {y_1}\,
      {y_2}}{4\, {t_2}\, {u_2} - 2\,{ {x_2}}^2}+\nonumber\\&&
    \frac{-\left(  {u_1}\,\left( -2\, {t_2}\,{ {x_2}}^2 +
         {u_2}\,\left( 2 + 2\,{ {t_1}}^2 + 2\,{ {t_2}}^2 +
           { {x_1}}^2 + { {x_2}}^2 \right)  \right) \,{ {y_2}}^2
      \right) }{2\,{\left( -2\, {t_2}\, {u_2} + { {x_2}}^2
        \right) }^2}+\nonumber\\&&
\frac{-\left( \left(  {x_1}\, {x_2}\,
         \left( -2\,{ {u_2}}^2 + { {x_2}}^2 \right)  -
         {t_1}\, {u_2}\,
         \left( 2 + 2\,{ {u_1}}^2 - 2\,{ {u_2}}^2 + { {x_1}}^2 +
           { {x_2}}^2 \right)  \right) \,{ {y_2}}^2 \right) }{2\,
    {\left( -2\, {t_2}\, {u_2} + { {x_2}}^2 \right) }^2}
   \nonumber\end{eqnarray}\begin{eqnarray}
      {\rm Im}({\Scr  N})_{0,4}&=&
- \frac{ {s_2}\,\left( { {x_1}}^3 -
        2\,\left(  {t_2}\, {u_1} +  {t_1}\, {u_2} \right)
           \, {x_2} +  {x_1}\,
         \left( 2 - 2\, {t_1}\, {u_1} +
           2\, {t_2}\, {u_2} + { {x_2}}^2 \right)  \right) }
      {{\sqrt{2}}\,\left( 2\, {t_2}\, {u_2} - { {x_2}}^2 \right)
        }+\nonumber\\&&
\frac{ {y_2}\,\left( -4\, {t_2}\,{ {u_2}}^2\, {x_2}\,
        {y_1} + 2\, {u_2}\,{ {x_2}}^3\, {y_1} +
      2\, {t_1}\,{ {u_2}}^2\, {x_2}\, {y_2} +
       {u_1}\,{ {x_2}}^3\, {y_2} \right) }{{\sqrt{2}}\,
    {\left( -2\, {t_2}\, {u_2} + { {x_2}}^2 \right)
    }^2}+\nonumber\\&&\frac{ {u_2}\, {x_1}\,
    \left( 2 - 2\, {t_1}\, {u_1} + { {x_1}}^2 -
      2\,{ {x_2}}^2 \right) \,{ {y_2}}^2}{{\sqrt{2}}\,
    {\left( -2\, {t_2}\, {u_2} + { {x_2}}^2 \right) }^2}
      \nonumber\\
      {\rm Im}({\Scr  N})_{0,5}&=&\frac{{\sqrt{2}}\,\left( -2\, {t_2}\,{ {u_2}}^2\, {y_1} +
      2\, {t_1}\,{ {u_2}}^2\, {y_2} +
       {u_1}\,{ {x_2}}^2\, {y_2} +
       {u_2}\, {x_2}\,
       \left(  {x_2}\, {y_1} - 2\, {x_1}\, {y_2} \right)
      \right) }{4\, {t_2}\, {u_2} - 2\,{ {x_2}}^2}\nonumber\\
      {\rm Im}({\Scr  N})_{1,1}&=& \frac{ {s_2}\,\left( { {t_1}}^2 + { {t_2}}^2 +
      2\, {t_1}\, {u_1} + { {u_1}}^2 + { {u_2}}^2 +
      { {x_2}}^2 \right) }{2\, {t_2}\, {u_2} -
    { {x_2}}^2}+\nonumber\\&&
    \frac{\left( \left( -{ {t_2}}^2 -
         {\left(  {t_1} +  {u_1} \right) }^2 \right) \, {u_2} +
      { {u_2}}^3 + \left(  {t_2} +  {u_2} \right) \,
       { {x_2}}^2 \right) \,{ {y_2}}^2}{{\left( -2\, {t_2}\,
         {u_2} + { {x_2}}^2 \right) }^2}
      \nonumber\\
      {\rm Im}({\Scr  N})_{1,2}&=&
- \frac{ {s_2}\,\left( -2\,
         \left(  {t_2} +  {u_2} \right) \, {x_1}\, {x_2}
         +  {u_1}\,\left( 2 + 2\,{ {t_1}}^2 + 2\,{ {t_2}}^2 -
           { {x_1}}^2 + { {x_2}}^2 \right)  \right) }{4\, {t_2}\,
        {u_2} - 2\,{ {x_2}}^2}+\nonumber\\&& - \frac{ {s_2}\, {t_1}\,
      \left( 2 + 2\,{ {u_1}}^2 + 2\,{ {u_2}}^2 - { {x_1}}^2 +
        { {x_2}}^2 \right) }{4\, {t_2}\, {u_2} -
      2\,{ {x_2}}^2}+\nonumber\\&&
\frac{ {y_2}\,\left( { {t_2}}^2\, {u_1}\, {u_2}\,
        {y_2} -  {t_1}\,{ {u_2}}^3\, {y_2} +
       {t_2}\,\left( 2\,{ {u_2}}^3\, {y_1} +
          {u_2}\,{ {x_2}}^2\, {y_1} -
          {u_1}\,{ {x_2}}^2\, {y_2} \right)  \right) }{{\left(
       -2\, {t_2}\, {u_2} + { {x_2}}^2 \right)
       }^2}+\nonumber\\&&
\frac{-\left(  {u_2}\, {y_2}\,
      \left( \left( - {t_1} -  {u_1} \right) \,
         \left( 2 + 2\, {t_1}\, {u_1} - { {x_1}}^2 -
           { {x_2}}^2 \right) \, {y_2} +
        2\, {u_2}\, {x_2}\,
         \left(  {x_2}\, {y_1} -  {x_1}\, {y_2} \right)
        \right)  \right) }{2\,{\left( -2\, {t_2}\, {u_2} +
        { {x_2}}^2 \right) }^2}+\nonumber\\&&
\frac{-\left( { {x_2}}^3\, {y_2}\,
      \left(  {x_2}\, {y_1} -  {x_1}\, {y_2} \right)
      \right) }{2\,{\left( -2\, {t_2}\, {u_2} + { {x_2}}^2
        \right) }^2}
      \nonumber\\
      {\rm Im}({\Scr  N})_{1,3}&=&\frac{ {s_2}\,\left( { {t_1}}^2 + { {t_2}}^2 -
       { {u_1}}^2 - { {u_2}}^2 \right) \,
     \left( 2\, {t_2}\, {u_2} - { {x_2}}^2 \right)  -
    \left(  {u_2}\,\left( { {t_1}}^2 + { {t_2}}^2 -
          { {u_1}}^2 + { {u_2}}^2 \right)  -
        {t_2}\,{ {x_2}}^2 \right) \,{ {y_2}}^2}{{\left( -2\,
         {t_2}\, {u_2} + { {x_2}}^2 \right) }^2}\nonumber\\
      {\rm Im}({\Scr  N})_{1,4}&=&- \frac{{\sqrt{2}}\, {s_2}\,
      \left( \left(  {t_1} +  {u_1} \right) \, {x_1} +
        \left(  {t_2} +  {u_2} \right) \, {x_2} \right) }{-2\,
        {t_2}\, {u_2} + { {x_2}}^2} +\nonumber\\&&
\frac{\left( -2\,\left(  {t_1} +  {u_1} \right) \, {u_2}\,
        {x_1} + 2\,{ {u_2}}^2\, {x_2} + { {x_2}}^3
      \right) \,{ {y_2}}^2}{{\sqrt{2}}\,
    {\left( -2\, {t_2}\, {u_2} + { {x_2}}^2 \right) }^2}
       \nonumber\\
      {\rm Im}({\Scr  N})_{1,5}&=&\frac{{\sqrt{2}}\,\left( 2\,{ {u_2}}^2 + { {x_2}}^2 \right) \,
     {y_2}}{4\, {t_2}\, {u_2} - 2\,{ {x_2}}^2}\nonumber\\
      {\rm Im}({\Scr  N})_{2,2}&=&\frac{ {s_2}\,\left( 4 + 4\,{ {t_1}}^2\,
       \left( { {u_1}}^2 + { {u_2}}^2 \right)  +
      4\,{ {t_2}}^2\,\left( { {u_1}}^2 + { {u_2}}^2 \right)  -
      4\,{ {x_1}}^2 + { {x_1}}^4 +
      2\,{ {x_1}}^2\,{ {x_2}}^2 + { {x_2}}^4 \right) }{8\,
      {t_2}\, {u_2} -
     4\,{ {x_2}}^2}+\nonumber\\&&
- \frac{ {s_2}\,\left(  {t_1}\,
         \left( 2\, {u_2}\, {x_1}\, {x_2} +
            {u_1}\,\left( -2 + { {x_1}}^2 - { {x_2}}^2 \right)
           \right)  +  {t_2}\,
         \left( 2\, {u_1}\, {x_1}\, {x_2} +
            {u_2}\,\left( -{ {x_1}}^2 + { {x_2}}^2 \right)
           \right)  \right) }{2\, {t_2}\, {u_2} - { {x_2}}^2}
    +\nonumber\\&&
\frac{ {y_1}\,\left( 2\, {t_2}\,{ {u_2}}^2\, {y_1} -
      4\, {t_1}\,{ {u_2}}^2\, {y_2} -
      2\, {u_1}\,{ {x_2}}^2\, {y_2} +
       {u_2}\, {x_2}\,
       \left( -\left(  {x_2}\, {y_1} \right)  +
         4\, {x_1}\, {y_2} \right)  \right) }{4\, {t_2}\,
      {u_2} - 2\,{ {x_2}}^2}+\nonumber\\&&
\frac{- {u_2}\,\left( 4\,{ {t_1}}^2\,
         \left(  {u_1} -  {u_2} \right) \,
         \left(  {u_1} +  {u_2} \right)  +
        4\,{ {t_2}}^2\,\left( { {u_1}}^2 + { {u_2}}^2 \right)  -
        (2+4\, {t_1}\, {u_1}-x_1^2)\,\left( -2 + { {x_1}}^2 \right)  \right)\,{ {y_2}}^2
    }{4\,{\left( -2\, {t_2}\, {u_2} + { {x_2}}^2 \right)
    }^2}+\nonumber\\&&
\frac{-  {x_2}\,\left( 8\, {t_1}\,{ {u_2}}^2\,
          {x_1} - 4\,\left(  {t_2}\,
            \left( { {u_1}}^2 + { {u_2}}^2 \right)  +
            {u_2}\,\left(  {t_1}\, {u_1} +
              { {x_1}}^2 \right)  \right) \, {x_2} +
        4\, {u_1}\, {x_1}\,{ {x_2}}^2 +
         {u_2}\,{ {x_2}}^3 \right) \,{ {y_2}}^2  }{4\,
    {\left( -2\, {t_2}\, {u_2} + { {x_2}}^2 \right) }^2}
     \nonumber\\
      {\rm Im}({\Scr  N})_{2,3}&=&- \frac{ {s_2}\,\left( 2\,{ {t_1}}^2\, {u_1} +
        2\,\left( - {t_2} +  {u_2} \right) \, {x_1}\,
          {x_2} +  {u_1}\,
         \left( -2 + 2\,{ {t_2}}^2 + { {x_1}}^2 - { {x_2}}^2
           \right)  +  {t_1}\,
         \left( 2 - 2\,{ {u_1}}^2 - 2\,{ {u_2}}^2 - { {x_1}}^2 +
           { {x_2}}^2 \right)  \right) }{4\, {t_2}\, {u_2} -
      2\,{ {x_2}}^2} +\nonumber\\&&
\frac{ {u_2}\,\left( 2\,{ {t_2}}^2\, {u_1} +
      2\, {t_1}\,{ {u_2}}^2 +
      \left(  {t_1} -  {u_1} \right) \,
       \left( 2 + 2\, {t_1}\, {u_1} - { {x_1}}^2 -
         { {x_2}}^2 \right)  \right) \,{ {y_2}}^2}{2\,
    {\left( -2\, {t_2}\, {u_2} + { {x_2}}^2 \right)
    }^2}+\nonumber\\&&
\frac{-\left(  {y_2}\,\left(  {x_2}\,
         \left( -2\,{ {u_2}}^2 + { {x_2}}^2 \right) \,
         \left(  {x_2}\, {y_1} -  {x_1}\, {y_2} \right)
         + 2\, {t_2}\,\left( 2\,{ {u_2}}^3\, {y_1} -
            {u_2}\,{ {x_2}}^2\, {y_1} +
            {u_1}\,{ {x_2}}^2\, {y_2} \right)  \right)  \right) }
    {2\,{\left( -2\, {t_2}\, {u_2} + { {x_2}}^2 \right) }^2}
      \nonumber\end{eqnarray}\begin{eqnarray}
 {\rm Im}({\Scr  N})_{2,4}&=&
\frac{ {s_2}\,\left( { {x_1}}^3 -
      2\,\left(  {t_2}\, {u_1} +  {t_1}\, {u_2} \right)
         \, {x_2} +  {x_1}\,
       \left( -2 - 2\, {t_1}\, {u_1} +
         2\, {t_2}\, {u_2} + { {x_2}}^2 \right)  \right) }{
    {\sqrt{2}}\,\left( 2\, {t_2}\, {u_2} - { {x_2}}^2 \right)
    }+\nonumber\\&&
- \frac{{\sqrt{2}}\, {u_2}\, {x_2}\, {y_2}\,
      \left( -2\, {t_2}\, {u_2}\, {y_1} +
        { {x_2}}^2\, {y_1} +
         {t_1}\, {u_2}\, {y_2} \right) }{{\left( -2\,
           {t_2}\, {u_2} + { {x_2}}^2 \right) }^2}
          +\nonumber\\&&
-\frac{\left(  {u_1}\,{ {x_2}}^3 +
         {u_2}\, {x_1}\,
         \left( -2 - 2\, {t_1}\, {u_1} + { {x_1}}^2 -
           2\,{ {x_2}}^2 \right)  \right) \,{ {y_2}}^2}{{\sqrt{2}}\,
      {\left( -2\, {t_2}\, {u_2} + { {x_2}}^2 \right) }^2}
   \nonumber\\
 {\rm Im}({\Scr  N})_{2,5}&=&\frac{{\sqrt{2}}\,\left( 2\, {t_2}\,{ {u_2}}^2\, {y_1} -
      2\, {t_1}\,{ {u_2}}^2\, {y_2} -
       {u_1}\,{ {x_2}}^2\, {y_2} +
       {u_2}\, {x_2}\,
       \left( -\left(  {x_2}\, {y_1} \right)  +
         2\, {x_1}\, {y_2} \right)  \right) }{4\, {t_2}\,
      {u_2} - 2\,{ {x_2}}^2}\nonumber\\
 {\rm Im}({\Scr  N})_{3,3}&=&\frac{ {s_2}\,\left( { {t_1}}^2 + { {t_2}}^2 -
      2\, {t_1}\, {u_1} + { {u_1}}^2 + { {u_2}}^2 -
      { {x_2}}^2 \right) }{2\, {t_2}\, {u_2} -
    { {x_2}}^2}+\nonumber\\&&
-\frac{\left(  {u_2}\,
         \left( {\left(  {t_1} -  {u_1} \right) }^2 +
           \left(  {t_2} -  {u_2} \right) \,
            \left(  {t_2} +  {u_2} \right)  \right)  +
        \left( - {t_2} +  {u_2} \right) \,{ {x_2}}^2 \right) \,
      { {y_2}}^2}{{\left( -2\, {t_2}\, {u_2} +
         { {x_2}}^2 \right) }^2}
    \nonumber\\
{\rm Im}({\Scr  N})_{3,4}&=&\frac{{\sqrt{2}}\, {s_2}\,\left(
 {t_1}\, {x_1} -
       {u_1}\, {x_1} +
      \left(  {t_2} -  {u_2} \right) \, {x_2} \right) }{2\,
      {t_2}\, {u_2} - { {x_2}}^2}+\nonumber\\&&
\frac{\left( -2\, {t_1}\, {u_2}\, {x_1} +
      2\, {u_1}\, {u_2}\, {x_1} -
      2\,{ {u_2}}^2\, {x_2} + { {x_2}}^3 \right) \,
    { {y_2}}^2}{{\sqrt{2}}\,
    {\left( -2\, {t_2}\, {u_2} + { {x_2}}^2 \right) }^2}
     \nonumber\\
     {\rm Im}({\Scr
N})_{3,5}&=&\frac{{\sqrt{2}}\,\left( -2\,{ {u_2}}^2 + { {x_2}}^2
\right) \,
     {y_2}}{4\, {t_2}\, {u_2} - 2\,{ {x_2}}^2}
    \nonumber\\{\rm Im}({\Scr  N})_{4,4}&=&\frac{ {s_2}\,\left( 2\, {t_2}\, {u_2} - { {x_2}}^2
       \right) \,\left( 2\, {t_2}\, {u_2} + 2\,{ {x_1}}^2 +
       { {x_2}}^2 \right)  +
    2\, {u_2}\,\left( - {x_1} +  {x_2} \right) \,
     \left(  {x_1} +  {x_2} \right) \,{ {y_2}}^2}{{\left( -2\,
         {t_2}\, {u_2} + { {x_2}}^2 \right) }^2}\nonumber\\
        {\rm Im}({\Scr  N})_{4,5}&=&\frac{2\, {u_2}\, {x_2}\, {y_2}}
  {2\, {t_2}\, {u_2} - { {x_2}}^2}\nonumber\\
  {\rm Im}({\Scr  N})_{5,5}&=&u_2
\end{eqnarray}}
%%%%%%%%%%%%%%%%%%%%%%%%%%%%%%%%%%%%%%%%%%%%%%%%%%%%%%%%%%%%%%%%%%%%%%%
%
%Use this if your figures are put in a subdirectory having the same
%name as the main latex file, ie:
%
%      ws-procs9x6/procs-fig1.eps
%      ws-procs9x6/procs-fig2.eps
%      ws-procs9x6/procs-fig3.eps
%      ws-procs9x6/procs-fig4.eps
%      etc.
%
%\begin{figure}[htbp] %ORIGINAL SIZE: width=1.4TRUEIN; height=1.5TRUEIN
%\figurebox{}{}{procf1} %100 percent
%\caption{Labeled tree {\it T}.}
%\end{figure}
%
%%%%%%%%%%%%%%%%%%%%%%%%%%%%%%%%%%%%%%%%%%%%%%%%%%%%%%%%%%%%%%%%%%%%%%%

\end{document}